\documentclass[10pt]{article}
\usepackage{graphicx}
\usepackage{geometry}
\usepackage{fullpage}
\usepackage{t1enc}
\usepackage{authblk}
\usepackage{enumitem}
\usepackage{color}
\usepackage{multicol}
\usepackage{ulem}
\usepackage{color}
\usepackage{tikz}
\usetikzlibrary{arrows,shapes,positioning,calc,decorations,math}
\usepackage{datenumber}
\usepackage{xifthen}
\usepackage{url}
\usepackage[symbol]{footmisc}

\newcounter{startdate}

\newcounter{tempdate}
\newcounter{dateone}
\newcounter{datetwo}

\newcommand{\datediff}[6]%
{   \setmydatenumber{dateone}{#1}{#2}{#3}
    \setmydatenumber{datetwo}{#4}{#5}{#6}
    \addtocounter{dateone}{-\thestartdate}
    \addtocounter{datetwo}{-\thestartdate}
}

\newcommand{\drawtimeline}%
{   \setdatebynumber{\thestartdate}
    \pgfmathtruncatemacro{\numberofdays}{\thetempdate}
    \pgfmathsetmacro{\daywidth}{\timelinewidth/\numberofdays}
    \draw[-stealth] (0,0) -- (\timelinewidth,0) -- ++(0.3,0);
    \foreach \x in {0,...,\numberofdays}
    { \ifthenelse{\thedateday = 1}
        { \ifcase\thedatemonth
            \or \xdef\monthname{Jan}
            \or \xdef\monthname{Feb}
            \or \xdef\monthname{Mar}
            \or \xdef\monthname{Apr}
            \or \xdef\monthname{May}
            \or \xdef\monthname{Jun}
            \or \xdef\monthname{Jul}
            \or \xdef\monthname{Aug}
            \or \xdef\monthname{Sep}
            \or \xdef\monthname{Oct}
            \or \xdef\monthname{Nov}
            \or \xdef\monthname{Dec}    
            \else       
            \fi
            \draw (\x*\daywidth,0) -- (\x*\daywidth,0.25) node[align=center,rotate=0,font=\small]
	    at (\x*\daywidth+15*\daywidth,0.25) {\monthname};
        }{}
        \ifthenelse{\equal{\datedayname}{Monday}}
        { \draw (\x*\daywidth,0) -- (\x*\daywidth,-0.05);
        }{}
        \addtocounter{datenumber}{1}
        \setdatebynumber{\thedatenumber}
    }
}

\newcommand{\timeentry}[9][gray]
{ \datediff{#2}{#3}{#4}{#5}{#6}{#7}
    \pgfmathtruncatemacro{\numberofdays}{\thetempdate}
    \pgfmathsetmacro{\daywidth}{\timelinewidth/\numberofdays}
    \draw[opacity=1,text opacity=1,line width=2.5mm,#1]
    (\thedateone*\daywidth,#8) -- (\thedatetwo*\daywidth,#8)
    node[left,rotate=0,font=\small,text opacity=1] at (0,#8) {#9};
}
\newcommand{\timeentryarr}[9][gray]
{ \datediff{#2}{#3}{#4}{#5}{#6}{#7}
    \pgfmathtruncatemacro{\numberofdays}{\thetempdate}
    \pgfmathsetmacro{\daywidth}{\timelinewidth/\numberofdays}
    \draw[opacity=1,text opacity=1,line width=2.5mm,>=fast cap,->,#1]
    (\thedateone*\daywidth,#8) -- (\thedatetwo*\daywidth,#8)
    node[left,rotate=0,font=\small,text opacity=1] at (0,#8) {#9};
}

\def\ind#1{$^{\mbox{\footnotesize{#1}}}$}

\def\TITLE{Symmetry energy at high densities from\\
neutron/proton flow excitation functions}

\begin{document}

\begin{titlepage}
\hspace{\fill} \textsc{June, 10, 2020}


\vspace{3mm}
	\vspace{2.0cm}

	{\centering
	{\scshape\large Proposal for Beam-time}\\
	
	{\scshape for constraining the}\\

	{\scshape\LARGE\bfseries \TITLE \par}
	}
	\vspace{1.0cm}

\hspace*{-5mm}
\begin{tabular}[h]{rl}
	{\scshape\Large Spokespersons:} & 
	{\Large\bfseries P. Russotto$^{*}$, A.~Le~F\`{e}vre, J.~\L{}ukasik } \\[2mm]
	&{\Large $^{*}$russotto@lns.infn.it}\\[2mm]
	 &{\Large INFN-LNS, Catania, Italy}\\[5mm] 
	
	
	{\scshape\Large Ion Species:} & {\Large\bfseries ${\bf ^{197}}$Au @ 250, 400, 600, 1000 AMeV}\\[5mm]


\end{tabular}

{\centering
{\scshape\Large Main Experimental/Scientific goals}\\[2mm]
}


{\large 

\begin{itemize}[leftmargin=*]
\item

{\bf Determination of the high density behavior of the symmetry energy} through
the simultaneous measurement of elliptic flow excitation functions of neutrons,
protons and light clusters. The elliptic flow developed in relativistic heavy
ion collisions has been proven theoretically and experimentally to have a unique
sensitivity and robustness in probing the symmetry energy up to around $2
\rho_{o}$. The knowledge of the density dependence of the symmetry energy in a
broad range of densities will provide a missing link for astrophysical
predictions of the neutron star mass--radius relation. In particular, the data
will provide  tighter constraints on the slope parameter L and entirely new
limits on $K_{sym}$, the currently poorly constrained symmetry energy curvature
parameter. 

\item

{\bf Enforcement of tight constraints on nuclear transport theories} by
providing new data on the symmetry energy and the inter-related phenomena of
clustering and neutron and proton emissions as well as correlations among them. 

\end{itemize}

\noindent The proposed campaign represents a unique set of measurements,
presently possible only at the GSI/FAIR facility, because of the available range
of beam energies and the existing instrumentation. 

}


	\vfill
\end{titlepage}

\newpage
\null
\thispagestyle{empty}


{\centering
{\scshape\large\bfseries \TITLE \par}
\vspace*{3mm}
}
\vspace*{5mm}
\noindent\textbf{\scshape  Spokespersons:} {P. Russotto\ind{1}, A.~Le~F\`{e}vre\ind{2}, J.~\L{}ukasik\ind{3}}
\vspace*{1mm}

\noindent\textbf{\scshape  Principal Investigators:} 
K.~Boretzky\ind{2}, M.D.~Cozma\ind{4}, E.~De~Filippo\ind{5}, I.~Ga\v{s}pari\'{c}\ind{6}, A.~Le~F\`{e}vre\ind{2}, Y.~Leifels\ind{2}, I.~Lihtar\ind{7}, J.~\L{}ukasik\ind{3}, S.~Pirrone\ind{5}, G.~Politi\ind{5,8}, P.~Russotto\ind{1}, W. Trautmann\ind{2} 
\vspace*{1mm}

\noindent\textbf{\scshape  Institutions:} 
\ind{1}INFN-LNS, Catania, Italy; 
\ind{2}GSI, Darmstadt, Germany; 
\ind{3}IFJ PAN, Krak\'{o}w, Poland; 
\ind{4}IFIN-HH, Bucharest, Romania;
\ind{5}INFN-Sezione di Catania, Italy;
\ind{6}RBI, Zagreb, Croatia; 
\ind{7}Univ. of Zagreb, Croatia;
\ind{8}Universit\`{a} di Catania, Italy;

 

\thispagestyle{empty}

\begin{abstract}

The proposed experimental program aims at putting new and more stringent
constraints on the density dependence of the symmetry energy at supra-saturation
densities. Densities toward $2\rho_{o}$, indispensable for obtaining the
constraints relevant for astrophysics, are expected to be reached in the
proposed central heavy ion collisions. {\bf \scshape Proposed systems}: Au+Au at
250, 400, 600 and 1000 AMeV. Such an energy scan can currently be
performed only at the GSI/FAIR facility. {\bf \scshape Key observables}:
excitation function of the neutron/proton elliptic-flow ratios, directed and
elliptic flows for n, p and isotopically resolved light clusters, yields and
ratios, energy and angular distributions, correlation functions. {\bf \scshape
The experimental setup} will be based on the NeuLAND detector for measuring
neutrons, protons and light charged clusters emitted from mid-rapidity. The main
novelty of this proposal consists in using NeuLAND to obtain well resolved
proton spectra allowing to probe effectively about 30\% higher densities than with only elemental resolution. The setup will include also the KRAB plastic barrel surrounding the target
for providing the multiplicity trigger and for estimating the centrality and the
reaction plane orientation by covering polar angles beyond $30^{\circ}$, four 
 double-rings of CHIMERA, and the R3B New Time-of-Flight Wall TOFD for extracting
event-by-event the centrality and the orientation of the reaction plane at polar angles up to
$30^{\circ}$. The START detector placed upstream of the target will provide the
reference signal for the time of flight measurement and trigger. The KRATTA
triple telescopes and the FARCOS array will be placed near the target for
measuring light charged particles at mid-rapidity and at backward angles,
respectively. {\bf \scshape Expected value added in nuclear and astrophysics}:
providing new constraints on the symmetry energy up to $2\rho_{o}$ from
simultaneous measurement of n, p and isotopically resolved light charged
particles.  Specifically, the new measurement is expected to provide tighter
constraints on the slope parameter L and entirely new ones on the, up to now
very weakly constrained, curvature parameter $K_{sym}$. Probing the densities
toward $2\rho_{o}$ is indispensable for meaningful comparisons with modern
nuclear theories. The new data on flow patterns and correlations, and on
competition between cluster and neutron and proton emissions will provide
valuable constraints for the transport models aiming at describing and
explaining these phenomena. Proposed measurement will provide results
complementary to those obtained with the ground and satellite based X-ray
telescopes and with the gravitational wave interferometers. The results are also
expected to be competitive in terms of precision. The knowledge of the Symmetry
Energy in a broad range of densities will provide a missing link for realistic
simulations of astrophysical objects and processes. {\bf \scshape Expected
instrumental value added}: commissioning of the KRAB and FARCOS detectors.

\end{abstract}

\vspace{3mm}

\setcounter{page}{1}

\noindent\textbf{\scshape Scientific Context and Motivation.}  The nuclear
matter Equation of State (EoS) is one of the central topics in contemporary
nuclear physics. In general, it describes the relation between density,
pressure, energy, temperature and the isospin asymmetry $\delta = (\rho_n -
\rho_p) / \rho$, where $\rho_n$, $\rho_p$, and $\rho$ are the neutron, proton
and nuclear matter densities, respectively. For cold nuclear matter it is
conventionally split into a symmetric matter part independent of $\delta$ and
an isospin term, expressed as a product of the symmetry energy\footnote{In the whole document we use the term "symmetry energy", instead of the more appropriate term "asymmetry energy", for consistency with what is commonly used by scientific community and in literature.}, $E_{sym}(\rho)$,
and $\delta^2$ \cite{ref:bao2008}: $ E(\rho, \delta)=E(\rho,
0)+E_{sym}(\rho)\delta^2+o(\delta^4)$.

Different density dependences of $E_{sym}(\rho)$ can be described 
quantitatively by expanding  $E_{sym}$ around the normal nuclear matter density,
$\rho_o$, leading to the following expression:

\begin{equation}
E_{sym}(\rho) = E_{sym,0}
+ \frac{L}{3}\left( \frac{\rho-\rho_o}{\rho_o} \right)
+ \frac{K_{sym}}{18}\left( \frac{\rho-\rho_o}{\rho_o} \right)^2 + ... 
\label{equ01-yl}
\end{equation}
\noindent where the value of the symmetry energy at normal density $E_{sym,0}
\equiv E_{sym} ( \rho = \rho_o)$, the slope parameter $L \equiv 3 \rho_o \;
\frac{\partial E_{sym}(\rho)}{\partial \rho} \Big |_{\rho = \rho_o}$, and the
curvature parameter (symmetry compressibility) $K_{sym} \equiv 9 \rho_{o}^{2}\;
\frac{\partial^{2} E_{sym}(\rho)}{\partial \rho^{2}}\Big |_{\rho = \rho_{o}}$. 

A theoretical determination of the nuclear EoS from first principles by
microscopic calculations is challenging and a subject of current scientific
research since several decades \cite{ref:fuchs2005}. In fact, microscopic calculations of the density functional of nuclear
matter employing different approaches to the nucleon-nucleon interaction predict
rather different forms of the EoS. In particular, the dependence of $E_{sym}$ on density $\rho$ shows very
different behaviors. 
Most calculations coincide at or slightly below normal nuclear matter density,
which demonstrates that constraints from finite nuclei are active for an average
density smaller than $\rho_o$ and surface effects play a role. In contrast to
that, extrapolations to supra-normal densities diverge dramatically, calling for
more tight experimental constraints in this region. However, significant
progress is currently being made: recently, calculations based on chiral
effective field theory ($\chi EFT$), combined with advanced statistical
methods,  have been able to predict the $E_{sym}$ at 2$\rho_{o}$ with about
5$\%$ precision \cite{ref:Dri20}. Nevertheless, heavy-ion laboratory experiments
and astrophysical measurements, see below, are needed to validate theoretical
findings.

The density dependence of the $E_{sym}$ is an important
constituent for the determination of the drip lines, masses, densities, and collective excitations of
neutron-rich nuclei \cite{ref:brown2010, ref:roca2011}, for flows and
multi-fragmentation in heavy-ion collisions at intermediate energies
\cite{ref:bao2008, ref:tsang2009}, but also for  astrophysical phenomena like
supernovae, neutrino emission, and neutron stars \cite{ref:steiner2005}, where
knowledge on the high-density dependence of the $E_{sym}$ is most
important.


In fact, one of the key question of modern physics is the determination of the
mass vs radius relationship of neutron stars. While the (maximum) mass of neutron
stars is mainly governed by the isoscalar nuclear matter equation of state,
$E(\rho,0)$, the radius of a neutron star is  governed by the symmetry energy
behavior at high density, around $2\rho_{0}$ \cite{ref:fat18}. In fact,
the pressure of neutron matter at $2\rho_{0}$ is what is basically needed to
determine the radius of a canonical neutron star; in \cite{ref:Lat16}, using a
multitude of EoS obtained by polytrope expansion, a very tight correlation
between pressure at $2\rho_{0}$ and radius of a 1.4 \(M_\odot\) neutron star was
obtained. A similar result was obtained in ref. \cite{ref:new09} using about 100
EoSes of different kind. In \cite{ref:sam18} masses and radii of neutron stars
were calculated from equations of state based on recent high-quality chiral
nucleon-nucleon potentials. For a 1.4 \(M_\odot\) neutron star predictions fall
between 10.8 and 12.8 km. Moreover it was shown that the radius of a 1.4
\(M_\odot\) neutron star is nearly insensitive to extrapolation beyond
$2\rho_{0}$. \\

\noindent \textit{The main aim of this proposal is to determine the Symmetry
Energy in the density region toward $2\rho_{0}$ which is relevant for realistic
simulations of astrophysical objects and processes.} \\

The discovery of gravitational waves has permitted a significant step-forward in
this field. In binary neutron star merger events, like the one observed in the
first evidence of gravitational wave  GW17082017 \cite{ref:Abb17}, one of the
key observable is the so called tidal polarizability $\Lambda$, strictly
correlated to the neutron stars radii. It follows that observation of GW
opened-up new opportunities for determining radii of neutron stars, possible
in the past only in few cases and with larger errors. Thus, the opening of
multi-messenger astronomy including gravitational waves  makes the study of the symmetry
energy at high density even more intriguing than in the past, allowing now a
direct and stringent comparison of data from terrestrial laboratories with astrophysical observations. A second event of  binary neutron
star merger GW190425 has been recently reported in \cite{ref:abb20}; there a
pressure of 19-80 $MeV/fm^{3}$ was estimated for neutron star matter at
$2\rho_{0}$.

A large amount of studies have been recently published, where the constraint on
symmetry energy from GW observation are compared with the ones early obtained in
terrestrial laboratory; thus, relevant step-forward arises from comparison of
results coming from nuclear and astro-physicists communities.  As an example, in
\cite{ref:zha19} Zhang and Li produced a restricted EoS parameter space using
observational constraints on the radius, maximum mass, tidal polarizability and
causality condition of neutron stars, resulting in an estimation of the Symmetry
Energy at $2\rho_{0}$ of $46.9\pm10.1~MeV$. It is interesting to note that in a
subsequent paper \cite{ref:zha19c}, the authors show that the observation of the
2.17 \(M_\odot\) neutron star reduces the error to $\pm 9 MeV$ and mention that
it is unlikely that even heavier neutron stars will be observed because the
value 2.17 \(M_\odot\) is already close to the theoretical maximum according to
several studies. In \cite{ref:xie19} a Bayesian analysis of GW170817 and
quiescent low-mass X-ray binaries radii suggested the symmetry energy at
$2\rho_{0}$ to be in the interval $31-51 MeV$. In \cite{ref:kra19, ref:zha19b}
the authors stated that while the tidal polarizability $\Lambda$ depends
strongly on the details of the symmetry energy, different trends of
$E_{sym}(\rho)$ lead to very similar values of $\Lambda$. Thus, measuring
$\Lambda$ alone may not determine completely the density dependence of the
symmetry energy; both nuclear laboratory experiments and astrophysical
observations are therefore necessary to break this degeneracy and determine
precisely the details of the symmetry energy. A similar conclusion comes from 
\cite{ref:for19} where it was shown how observations of gravitational waves from
binary neutron star mergers can be combined with insights from nuclear physics
to obtain useful constraints on EoS of dense matter between one and two times
the nuclear saturation density. Moreover, first results for the radius of a 1.4 \(M_\odot\) neutron
star from X-ray pulse-profile modeling have been reported by the NICER (Neutron star Interior Composition Explorer) collaboration very recently \cite{ref:ril19, ref:mil19}. It will be interesting to see the impact of the comparably large values of 12.7 km or 13.0 km with errors of $\pm$ 1.2 km in comparison with the data from other sources.\\
In the last two decades,  meaningful constraints for the nuclear EoS  have been
obtained by laboratory experiments. Many results of nuclear structure and
nuclear reaction measurements as well as astrophysical observations were
collected and compared in  \cite{ref:bao2013} and \cite{ref:lattimer2014}.
Rather precise values of $E_{sym}$ have been evaluated for $\rho/\rho_o \sim
0.6-0.7$ in \cite{ref:brown2013} and \cite{ref:zhang2013} by fitting the
properties of doubly closed-shell nuclei.  Together with the results of an
analysis of isobaric analogue states \cite{ref:danielewicz2014} and from heavy
ion reaction data \cite{ref:tsang2009}, one obtains a quite consistent behavior
of $E_{sym}$ at low densities \cite{ref:horowitz2014}. 

As mentioned above, the $E_{sym}$ above $\rho_{0}$ can be accessed either by the
determination of the masses and radii of neutron stars \cite{ref:lattimer2014}
or by employing observables in heavy ion collisions which are related to the
early, high density phase of the reactions. A multitude of observables have been
proposed to be sensitive to the $E_{sym}$ at supra-saturation densities (for a
review see \cite{ref:bao2008}): ratio of multiplicities or spectra of isospin
partners (e.g. $\pi^-/\pi^+$, n/p or t/$^3$He) and the comparison of their
flows. The ratio of positively and negatively charged pions, as measured in the 1 AGeV regime for various collision systems by the FOPI collaboration \cite{ref:reisdorf2007}, were well reproduced with the IBUU4
transport code \cite{ref:xiao} but only with a super-soft density dependence of $E_{sym}$.
The incorporation of in-medium effects like mass-shifts of the pions, pion potentials, s-wave
production of pions, and the properties of intermediate $\Delta$ resonances has been shown to
lead to different and even opposite conclusions \cite{ref:feng,ref:danielewicz}, while describing the experimental data equally well, indicating a strong model dependence in the interpretation of the pion ratio. 
As a solution, it has been proposed to study not the integrated pion yield ratios but
the ratios of pion spectra at high kinetic energies in the
center of mass (CM) reference frame \cite{ref:tsang2017}.
Experiments with this aim have been carried out at RIKEN with the SAMURAI magnet and the SPIRIT TPC in 2016
\cite{ref:Spirit}, and $\pi^-/\pi^+$ ratios in
neutron rich and neutron poor Sn+Sn collision have been measured at 270 AMeV. The sensitivity to the
symmetry energy is high around the pion production threshold but the range of densities effectively probed will be centered below saturation density \cite{ref:yon19}.
In parallel to this experimental activity, a strong theoretical
effort is being made, comparing pion predictions from
several models within the code comparison project aiming to better
understand and eventually reduce systematic differences between model predictions \cite{ref:Xu16,ref:Zha18,ref:Ono19}. The aim is to support the interpretation of this new set of pion data.

Other observables which are known to be sensitive to $E_{sym}$ at supra-normal
densities are collective flows. At energies below 1.5A~GeV the reaction
dynamics is largely determined by the nuclear mean field. The resulting pressure
produces a collective motion of the compressed material whose strength will be
influenced by $E_{sym}$ in isospin-asymmetric systems.  The strengths of
collective flows in heavy ion collisions are usually expressed with a Fourier expansion
of the azimuthal distributions of particles around the reaction plane: 
$d\sigma(y)/d\phi \propto 1 + 2(v_1(y) \cos{\phi} + v_2(y) \cos{2\phi}.....)$.
The side flow of particles is characterized by the coefficient $v_1$ and the
elliptic flow by $v_2$. The value of $v_2$ around mid-rapidity is negative at
incident beam energies between 0.2 and 5~AGeV which signifies that matter is
squeezed out perpendicular to the reaction plane.  At intermediate energies
(below 1.5A~GeV), the elliptic flow of protons and composite charged particles emitted at
midrapidity in heavy-ion collisions shows the strongest sensitivity to the
nuclear equation of state (EoS) \cite{ref:lefevre2}. Thus, $v_2$ is a strong tool for constraining the strength of the mean field, of both isoscalar and isovector contributions.  In fact, as an example, the EoS of symmetric matter was strongly constrained in \cite{ref:lefevre} by the elliptic flow $v_2$ of
protons, deuterons, tritons, $^{3}He$ and $^{4}He$ emitted around mid-rapidity
in Au+Au collisions \cite{ref:reisd2012}. This study concluded in favor of a
soft momentum dependent EoS when comparing with the IQMD model predictions
\cite{ref:hartnack}. This result has been  more recently confirmed in a similar
comparison, using UrQMD transport model calculations \cite{ref:wang18}. 
For the case of asymmetric matter, the ratio $v_{2}^n/v_2^{p(ch)}$ of the
elliptic flow strengths of neutrons with respect to that of protons or light charged particles 
was recommended in \cite{ref:russotto2011} as a robust observable sensitive
to the stiffness of the $E_{sym}$.\\

\begin{figure}[ht!]
 \begin{minipage}[t]{0.5\linewidth}
 \begin{center}
\includegraphics[width=0.75\textwidth]{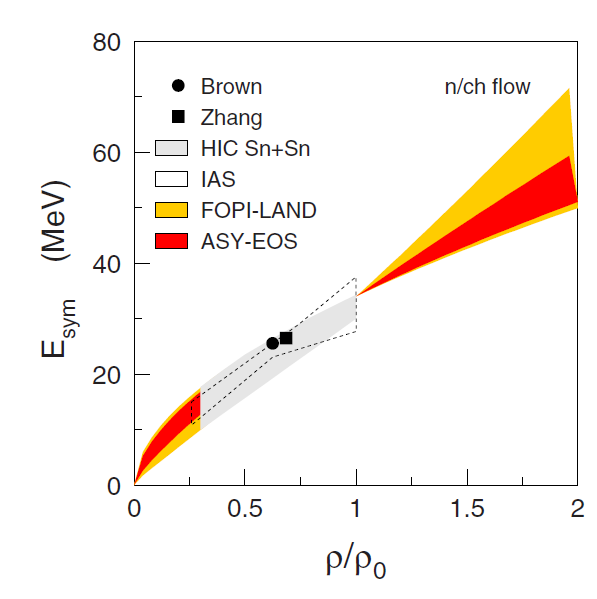}
 \end{center}
 \end{minipage}
 \hfill
 \begin{minipage}[t]{0.5\linewidth}
    \vspace*{-65mm}
 \begin{center}
\includegraphics[width=0.98\textwidth]{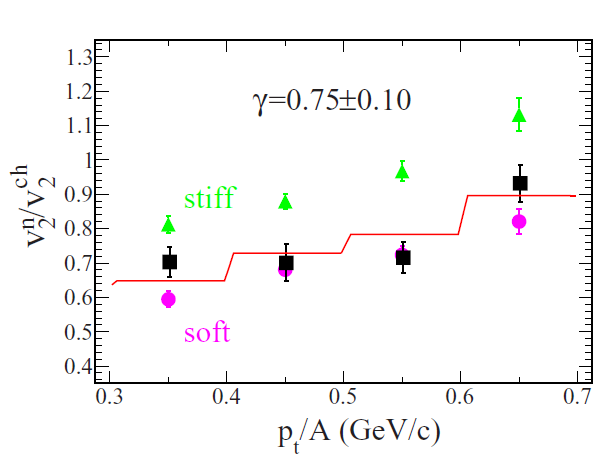} 
 \end{center}
 \end{minipage}
\vspace{-7mm}

\caption{ Left panel: constraints deduced for the density dependence of the
symmetry energy from the ASY-EOS (orange band) and FOPI-LAND (yellow band) experiments, compared also to some low-density results. Right panel: Elliptic flow
ratio of neutrons over charged particles measured in the same acceptance range
for central ($b < 7.5$~fm) Au+Au collisions at 400~AMeV as a function of
transverse momentum,  $p_t/A$. The black squares represent the ASY-EOS
experimental data. The green triangles and purple circles represent the UrQMD results employing
a stiff and soft density dependence of the $E_{sym}$. The solid line is the result of a linear interpolation between the predictions leading to the indicated  $\gamma=0.75\pm0.10$. From ref. \cite{ref:russotto2016}.}

\label{fig:fig3}
\end{figure}


\noindent\textbf{\scshape Previous Experiments and Background.} The first
attempt to constrain the $E_{sym}$ at high densities has been done in
\cite{ref:russotto2011} by re-analyzing the existing FOPI-LAND data
\cite{ref:leifels1993} on neutron and Hydrogen emissions obtained with the LAND
\cite{ref:blaich} detector. The re-analysis using the $v_{2}^n/v_2^{H}$ ratio
and comparison to the results of the UrQMD transport model \cite{ref:li} yielded
a moderately soft $E_{sym}$ dependence on density, with the slope parameter $L =
83 \pm 26$~MeV. Despite a large uncertainty, the result made it possible to rule
out the extremely soft or stiff density dependencies of the symmetry energy
\cite{ref:coz13}. It was, in particular, also possible to demonstrate that the
effects of isoscalar-type parameters affecting the individual flows in the
calculations largely cancel in the predictions for the flow ratios. 

Motivated by this finding, an attempt was made to improve the accuracy with a
new experiment that was conducted at the GSI laboratory in 2011 (ASY-EOS
experiment S394 \cite{ref:russotto2016}). The experimental setup followed the scheme developed for
FOPI-LAND by using LAND as the main instrument for neutron and charged particle
detection. For the event characterization and for measuring the orientation of
the reaction plane, three detection systems had been installed, the ALADIN
Time-of-Flight (AToF)Wall \cite{ref:sch96},  four double rings of the CHIMERA
multidetector \cite{ref:chimera} carrying together 352 CsI(Tl) scintillators in
forward direction and four rings with 50 thin CsI(Tl) elements of the Washington
University Microball array \cite{ref:sar96} surrounding the target. A detailed
description of the set-up of the ASY-EOS experiment and of the data analysis
procedure is available in \cite{ref:russotto2016}. Constraints on the symmetry
energy were obtained by comparing the experimental $v_{2}^n/v_2^{ch}$ ratios,
neutron over charged-particles, with those from the UrQMD simulations. Being not
able for technical reasons to well identify protons has been one of the main
drawback of that measurement, reducing also the maximum density effectively
probed. A soft iso-scalar EoS was assumed for the simulations and the $E_{sym}$
was parametrized with a Fermi-gas-like formula: 
\begin{equation}
E_{sym} (\rho) =
E_{sym}^{pot}(\rho) + E_{sym}^{kin}(\rho) =  22 (\rho/\rho_o)^{\gamma} +
12(\rho/\rho_o)^{2/3}\; \mbox{MeV} 
\label{equ02}
\end{equation}
\noindent with $\gamma$ = 0.5 and $\gamma$ = 1.5 corresponding to a soft  and a
stiff density dependence,   respectively.

\begin{figure} [h]
\centering
\includegraphics[width=0.29\textwidth]{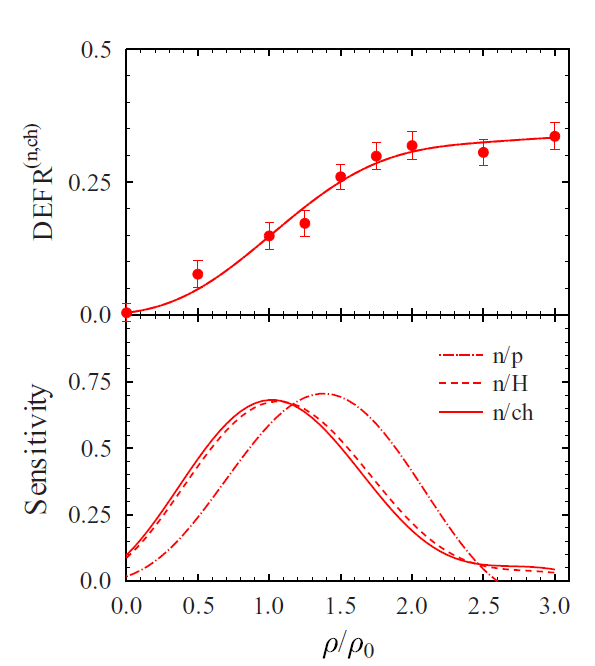}
\includegraphics[width=0.33\textwidth]{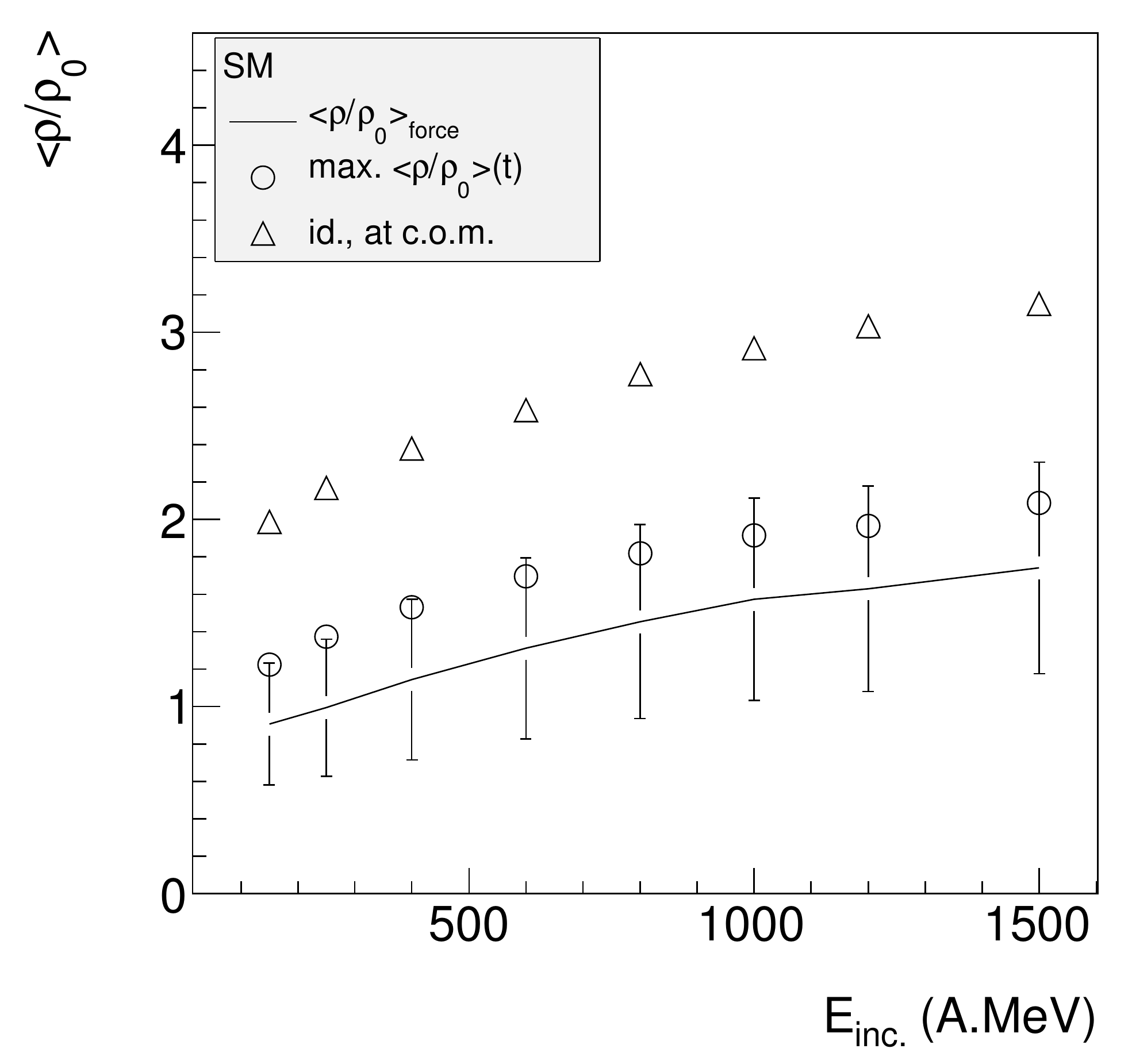} 

\caption{Left panel: (top) T\"uQMD predictions of the $DEFR$ function for the
ratio of elliptic flows of neutrons and charged particles for the Au+Au system
at 400A~MeV; (bottom) corresponding sensitivity density (solid line) together
with the ones obtained from elliptic-flow ratios of neutrons over all hydrogen
isotopes (dashed line) and neutrons over protons (dash-dotted line). From Ref.
\cite{ref:russotto2016}. Right panel: IQMD predictions of the incident energy
dependence of the average reduced density $<\rho/\rho_0>$ of protons in
semi-central (impact parameter $b=3 fm$) collisions of $^{197}$Au+$^{197}$Au
(with the soft momentum-dependent EoS "SM" which best reproduced the
experimental data), for various space-time selections: (triangles) maximum value
reached in the central volume of the collision, (circles) maximum value for
protons ending-up in the phase space selected for constraining the EoS in
\cite{ref:lefevre} (reduced transverse velocity ut0 > 0.4 and reduced rapidity
|y0| < 0.8), (error bars) spread distribution (one sigma) of the time averaged
value weighted by the force of the mean field felt by protons falling in the
same phase space selection. See \cite{ref:lefevre} for more details.}

\label{fig:fig2}
\end{figure}

The main results of the experiment are shown in Fig.~\ref{fig:fig3}. From the
fit of the measured flow ratios with a linear interpolation between the soft and
stiff model predictions the exponent {$\gamma = 0.75 \pm 0.10$} was obtained
(right panel of Fig.~\ref{fig:fig3}). After taking into account all corrections
and systematic uncertainties, the final value was found to be $\gamma = 0.72 \pm
0.19$, corresponding to a slope parameter $L = 72 \pm 13$ MeV
\cite{ref:russotto2016}. The corresponding density dependence of the Symmetry
Energy is shown in left panel of Fig. ~\ref{fig:fig3}. It confirms the FOPI-LAND
result and represents an improvement of the accuracy by a factor of two. The
sensitivity of the measurement to the density that is probed with the flow ratio
in the studied reaction is shown in Fig. ~\ref{fig:fig2}. It is centered at
approximately saturation density and extends beyond twice that value. The
sensitivity distribution indicates that the flow ratio  $v_{2}^n/v_2^{ch}$ at this bombarding energy
is well suited to measure the density dependence of the symmetry energy slightly
above the saturation density $\rho_{0}$.

The ASY-EOS constraints suggest the 51-60 MeV interval for the $E_{sym}$ at 
$2\rho_{0}$ in partial overlap with above cited values of the Refs.
\cite{ref:zha19,ref:xie19}. The obtained slope parameter $L = 72 \pm 13$ MeV,
corresponding to a symmetry pressure $p_{0} = 3.8 \pm 0.7$ $MeV fm^{-3}$, was
used to estimate the pressure in neutron star matter at density $\rho_{0}$
\cite{ref:russotto2016}. The obtained value $3.4 \pm 0.6$ $MeV fm^{-3}$ is
located inside the pressure interval obtained with $95\%$ confidence limit by
Steiner et al. from the observation of eight neutron-stars \cite{ref:ste13} and
inside the $90\%$ confidence interval of the pressure-density relation presented
by the LIGO and Virgo collaborations \cite{ref:abb18}. The known tight
correlation between the symmetry pressure $p_{0}$ and the radius $R_{1.4}$ of a
canonical neutron star of 1.4 \(M_\odot\) \cite{ref:Lat16} has, furthermore,
been used and a value $R_{1.4} = 12.6 \pm 0.7$ km was obtained \cite{ref:tra19}.
It is in amazingly good agreement with the recently published radii $12.7 \pm
1.2$ km \cite{ref:ril19} and $13.0 \pm 1.2$ km \cite{ref:mil19} of the NICER
collaboration, and the errors are very competitive. The ASY-EOS $R_{1.4}$
estimation agrees also with the values of Ref. \cite{ref:sam18} above reported.
Since both GW and X-rays direct observation are in a starting phase, and ASY-EOS
estimations in the $2\rho_{0}$ region should be taken as an extrapolation, these
agreements should be considered as preliminary. 

The comparisons illustrate the value of terrestrial measurements of the
EoS of asymmetric matter to complement and possibly support astrophysical
observations whose uncertainties are related to the complexity of the applied
methods \cite{ref:Xie20} or to serve as starting points for extrapolations to high densities exceeding $2 \rho_{0}$ \cite{ref:Bao20}. It has to be emphasized, however, that the UrQMD analysis of the
ASY-EOS flow ratios relies on two assumptions. The expression for $E_{sym}$ 
(Eq. \ref{equ02}) assumes $E_{sym}(\rho_{0}) = 34$ MeV, leading to the sharp
cross over of the error bands visible in Fig.~\ref{fig:fig3}. It does not
reflect the present uncertainty of approximately 3 MeV of our knowledge of the
symmetry energy at saturation \cite{ref:bao2013,ref:oer17}. $E_{sym}(\rho_{0}) =
31$ MeV in the analysis reduces the resulting slope parameter to $L = 63 \pm 11$
MeV as reported in the ASY-EOS publication \cite{ref:russotto2016}.  The second
assumption is that of the functional form of a power law (Eq. \ref{equ02}) that,
with the present results, is equivalent to assuming -70 MeV to -40 MeV for
$K_{sym}$, an interval that does not at all correspond to our limited knowledge
of the curvature term. However, as shown by Cozma \cite{ref:coz18}, these
difficulties can be overcome with measurements performed at different energies
and by exploiting the dependence of the sensitivity to density on the type of
charged particles whose flow is selected for the comparison with the neutron
flow (Fig.~\ref{fig:fig2}). The proposed measurements are intended to serve that
purpose. Thus, a new experiment probing with high effectiveness the region
toward $2\rho_{0}$ will allow to get better and more reliable constraints from
terrestrial laboratories to be compared with new and more systematic data coming
form astrophysical observations.

In \cite{ref:coz18}, Cozma reported a more complete analysis of FOPI flow data,
FOPI-LAND and ASY-EOS by using a QMD type transport model supplemented by a
phase-space coalescence model fitted to FOPI experimental multiplicities of free
nucleons and light clusters. Considering that  calculation has proven that
neutron-to-proton and neutron-to-charged particles elliptic flow ratios probe on
average different densities, see below, and anchoring symmetry energy
parametrization at the precise value available for $\rho=0.1 fm^{-3}$, Cozma
extracted both the slope L and curvature $K_{sym}$ parameters of the Symmetry
Energy, as $L=85\pm22(exp)\pm20(th)\pm12(sys)$ MeV and
$K_{sym}=96\pm315(exp)\pm170(th)\pm166(sys)$ MeV.  Measuring at the same time
yields and flow of LCP, as we plan to do, is a way to reduce the systematic
error arising from light-cluster multiplicities not well reproduced by the
models. Also a value of L, free of systematical theoretical uncertainties, was
extracted from the neutron-to-proton elliptic flow ratio alone,
$L=84\pm30(exp)\pm19(th)$ MeV.


For the Au+Au case at 400A~MeV,  the specific density region  tested by the flow
ratio observable has been explored in \cite{ref:russotto2016} within the T\"uQMD
model \cite{ref:cozma2011aa}. Two calculations have been performed there using a
moderately soft and a stiff $E_{sym}$ dependencies up to a given density
threshold and using a common intermediate dependence on density above that
threshold. The Difference of Elliptic Flow Ratios, $DEFR$, with these two
parametrizations has been then defined as an ``observable'' measuring the
sensitivity of the density region below the threshold on the stiffness of the
$E_{sym}$. The left top panel of Fig.~\ref{fig:fig2} shows the $DEFR$ function
for the $v_2$ ratios of neutrons to charged particles. It can be seen that the
sensitivity achieved with the elliptic flow ratios increases with the density
threshold and then saturates going toward the $2\rho_{0}$ region. To evaluate
this more carefully,  bottom panel shows the derivative of the  $DEFR$ (solid
line), to be used as a quantitative estimator of the sensitivity of the v2-ratio
observable to a given density region. Dashed and dash-dotted line report the
same quantity but for v2-ratio of neutrons with respect to Hydrogen's or
protons, respectively.  For the Au+Au case at 400 AMeV, v2-ratio of neutrons
with respect to charged-particles or Hydrogen's is mainly sensitive to a region
centered slightly above saturation density, while the n/p ratio sensitivity is
centered around 1.4$\rho_{0}$.\\

\noindent \textit{Thus, the direct measurement of n/p ratio, instead of
n/charged particles ratio, will enable more effective probing of much higher density
region than was possible so far. This will be a very big step forward.}\\


This observation gives rise to the expectation that with an isotopic resolution
sufficient for unambiguous proton identification one should be able not only to
constrain the slope of the $E_{sym}$, $L$, but also its curvature, $K_{sym}$,
see eq. (\ref{equ01-yl}). The latter is the least constrained experimentally and
theoretically observable so far. The experimental constraint of
\cite{Centelles2009}  yields  $K_{sym}=-50\pm$200 MeV. Another compilation
\cite{bao} locates the theoretical value of $K_{sym}$ between $-400$ and +466
MeV and its experimental value in a range from $-566\pm$1350 MeV to +34$\pm$159
MeV. The results of Cozma \cite{ref:coz18} are given above. In Fig. 13 of this
work it was also shown that for Au+Au semi-central collisions neutron-proton
elliptic flow ratio sensitivities to L and $K_{sym}$ attain maxima at 600  and
250 AMeV, respectively, thus at quite different energies. Note that the
potential terms that are proportional to L and $K_{sym}$ have different
dependencies on density and, consequently, the forces generated by these two terms
attain their maximum effectiveness at different regions of density; simplifying, the L term is
proportional to the isospin asymmetry $\delta$, while the $K_{sym}$ term to
$\delta\rho$. This explains why the maximum sensitivity for L and $K_{sym}$ do
not occur for the same incident energies. This is another strong reason for
measuring excitation function of neutron-proton elliptic flow observable.

\begin{figure}[h]
\centering
\includegraphics[width=1.\textwidth]{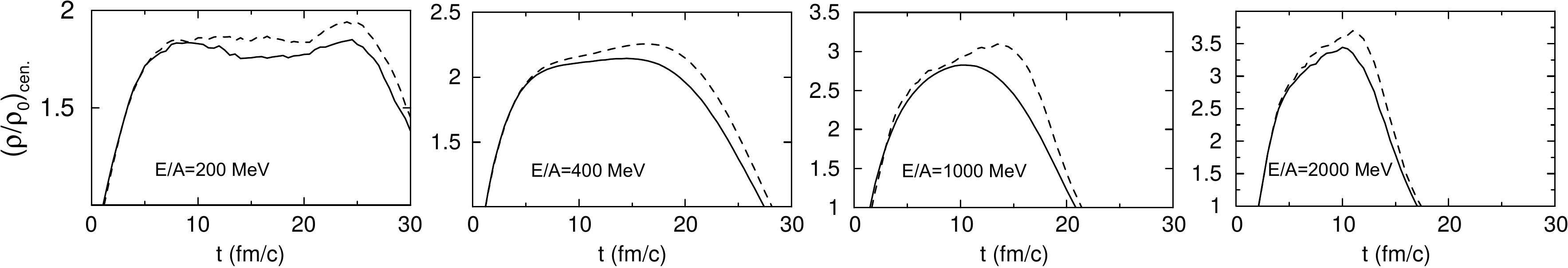} 

\caption{Evolution of the central baryon density in $^{132}$Sn+$^{124}$Sn
collisions at beam energies from 200 to 2000 AMeV for b = 1 fm, as  predicted by the
hadronic transport model of \cite{ref:Bao-An02}. }

\label{fig:fig6}
\end{figure}

Microscopic transport calculations predict that for a short time period
($\sim$20~fm/c) densities of up to $\sim$3 times the saturation density can be
reached in the central zone of a heavy-ion collision even at moderate incident
energies $\sim$1A~GeV, as demonstrated in Fig.~\ref{fig:fig6}.  As an example,
the right panel of Fig.~\ref{fig:fig2} provides also convincing arguments that
the maximum densities in the innermost center of the collision may reach up to
3.5$\rho_0$ at 1.5 A~GeV, while the densities probed by the mid-rapidity protons
during the heavy ion collision may extend up to twice the saturation density,
irrespectively of the transverse momentum cut. This finding comes from
\cite{ref:lefevre} and was obtained by analyzing the elliptic flow data of
charged particles and employing the Quantum Molecular Dynamics code IQMD
\cite{ref:hartnack} for Au+Au collisions up to 1.5~AGeV. Analogously to the
above mentioned DERF function used with T\"uQMD calculations,  this study has
given IQMD predictions of ranges of densities that are probed by the effect of
the mean field in the  elliptic flow, that is the "force-weighted" average
density shown as error bars in the right panel of Fig.~\ref{fig:fig2}.  It shows
that for Au+Au collisions at 1A~GeV, the typical densities directly influencing
the flow by the way of the mean field span between $\rho_0$  and 2.2$\rho_0$.
Nevertheless, as demonstrated in \cite{ref:lefevre2}, higher densities reached
in the compressed central region of the colliding system (dubbed as "fireball")
play also an indirect role in the strength of the elliptic flow, because they
determine how fast the later expansion of the fireball will occur and modify the
elliptic flow by the interplay between expanding fireball and flying away
spectators. And the speed of this expansion is ruled by the strength of the mean
field. Therefore, at 1A~GeV, the influence of the EoS at 3$\rho_0$ on $v_2$
cannot be ruled out. This conclusion applies also to the isovector part of the
EoS. In the most compressed phase of the collision, it is expected that due to
the symmetry energy, the neutron part in excess of the fireball will expand
faster than protons, then will interact differently with the spectators, which
will result in a different elliptic flow.  

Therefore, according to the models, it is possible to access the $E_{sym}$ at
higher densities by raising the beam energy. However, as the incident energy is
increased the fraction of nucleons excited into baryonic resonances, mainly
$\Delta(1232)$ at energies of 1 AGeV and below, in the highly compressed phase
of the collision reaches values in the neighborhood of $20\%$ \cite{ref:PRC47}
with potential impact on the time evolution of the reaction that may leave a
comparable imprint on the $E_{sym}$ at 2-3 $\rho_{0}$ depending on the chosen
observables. Thus, one should be aware that the highest density reached during a
heavy-ion collision is not necessarily equivalent to the density that can safely
be probed, i.e. without the occurrence of unmanageable systematic theoretical
uncertainties, for the purpose of constraining the high density dependence of
the $E_{sym}$, as discussed  in \cite{ref:russotto2016}, \cite{ref:lefevre} and
\cite{ref:lefevre2}. The ASY-EOS II measurements will provide powerful data to
explore these delicate aspects.\\

\noindent\textbf{\scshape New Approach and Relevance to the Field.} The main
novelty of the proposal consists in using the NeuLAND detector for measuring
neutrons, isotopically resolved H and He isotopes in a broad energy range within
the same acceptance. This will be the first measurement of this kind and
quality, allowing the high density $E_{sym}$ to be pinned down with an
unprecedented precision. Using the 12 double planes of NeuLAND to obtain
isotopically well resolved proton spectra will allow us to effectively probe about
30\% higher densities than was possible so far. New detectors, KRAB and FARCOS,
will improve the precision and quality of the data as detailed below. Measuring
excitation functions of flow observables will provide additional constraints on
$E_{sym}$ by scanning through a broad range of densities.

New tight constraints on the symmetry energy for densities reaching $2\rho_{o}$
will complement the results obtained with the X-ray telescopes and with the GW
interferometers and are also expected to be competitive in terms of precision.
The knowledge of the symmetry energy in a broad range of densities will provide
the missing information for astrophysical predictions of the neutron star mass--radius
relation and for realistic simulations of neutron stars, supernova explosions
and nucleosynthesis.

Precise data on the symmetry energy and the inter-related phenomena of
clustering and neutron and proton emissions as well as correlations between them
will present strong constraints to nuclear transport theories. The data should
allow theorists to address problems such as delta and pion production and dynamics,
importance of three-body forces at high densities, cluster formation, and effective
neutron and proton mass splitting related to the momentum dependence of
the nuclear mean field. \\



\begin{figure}[]
\centering
\includegraphics[width=0.475\textwidth]{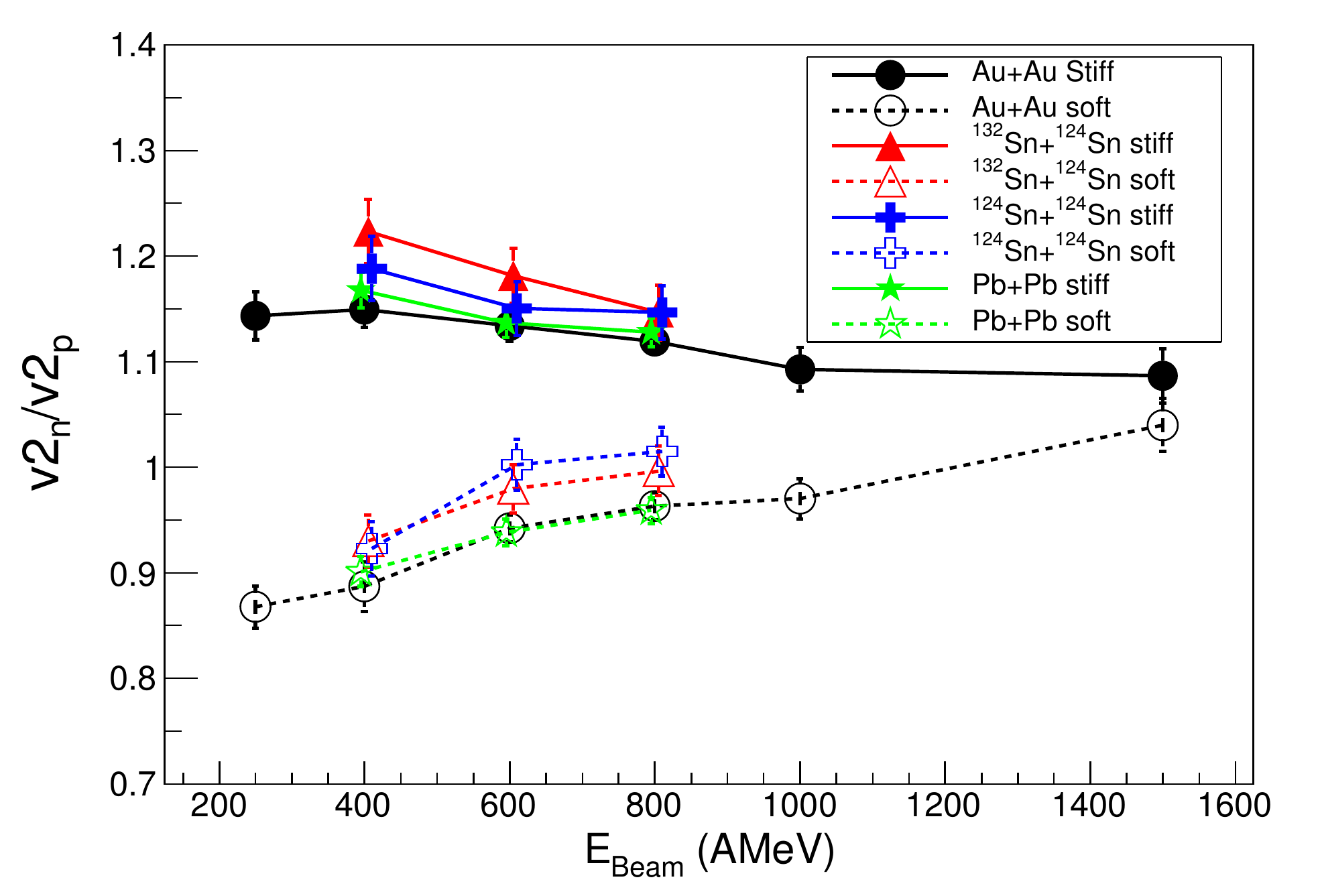}
\includegraphics[width=0.475\textwidth]{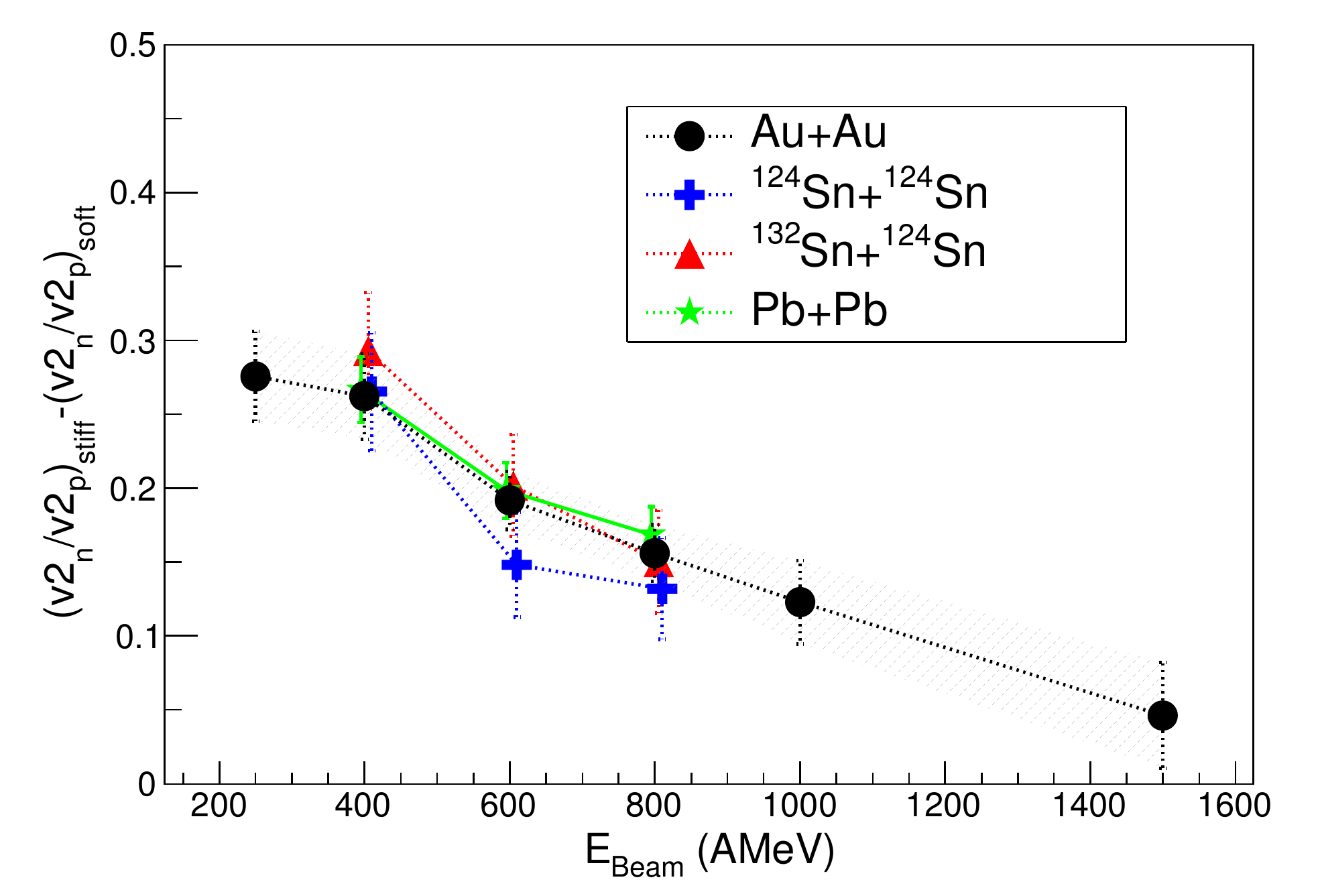}

\caption{Left panel: Excitation functions of neutron-to-proton elliptic flow
ratios, $v2_{n}/v2_{p}$, at mid-rapidity for semi-central Au+Au,
$^{132}$Sn+$^{124}$Sn, $^{124}$Sn+$^{24}$Sn and Pb+Pb collisions, as predicted by the
UrQMD model for stiff and soft $E_{sym}(\rho)$. Right panel: differences between
the stiff and soft results. }

\label{fig:fig7}
\end{figure}

\noindent\textbf{\scshape Objectives, Expected Results and Theory Background.}
The ASY-EOS experimental results proved the effectiveness of the
$v_{2}^n/v_2^{ch}$ ratio in constraining the high-density behavior of the
$E_{sym}$. The method appears to be very robust and precise; we can notice that
statistical errors smaller than $10\%$ can be obtained. As seen in
Fig.~\ref{fig:fig2} (right panel), \textit{ although the accurate ASY-EOS
determination of the $E_{sym}$ is estimated to have reached the supra-saturation
density region, which was already a unique achievement, there remains a strong
need for constraining it further, with an even better precision given by the
more effective $v_{2}^{n}/v_{2}^{p}$ ratio, toward $\sim$2 $\rho_{0}$. SIS@GSI provides a unique tool to probe such densities with heavy-ion
collisions.} Simulations of semi-central Au+Au collisions at 250, 400, 600, 800,
1000 and 1500 AMeV and, for comparison, neutron rich
$^{132}$Sn+$^{124}$Sn, $^{124}$Sn+$^{124}$Sn and Pb+Pb  systems at 400, 600 and
800 AMeV have been carried out by using the same version of the UrQMD
transport model already used in \cite{ref:russotto2016}.  The neutron-to-proton
elliptic flow ratio, $v_{2}^{n}/v_{2}^{p}$, at mid-rapidity
($0.4<y_{lab}/y_{proj}<0.6$), with a stiff ($\gamma$=1.5) and a soft
($\gamma$=0.5) parametrization of the potential part of the $E_{sym}$ for
semi-central $(b_{red}<0.54)$ collisions is shown, as a function of the incident
beam energy, in the left panel of Fig.~\ref{fig:fig7}. The difference of such
$v_{2}^{n}/v_{2}^{p}$ ratios between the stiff and soft choices can be taken as the
sensitivity of the proposed observable, and is shown in the right panel of the
same figure.

The obtained sensitivity  decreases with increasing the beam energy, 
because the mean-field
contribution decreases at higher energies where the two-body collisions start to
dominate. Nevertheless, up to 1 AGeV the sensitivity  of the proposed observable
is $\sim$15\%, while a measurement can easily reach a $\sim$5\% accuracy,
allowing clear discrimination between stiff and soft choices. 

A similar conclusion can be drawn from  a recent paper \cite{ref:wan20} where
UrQMD simulation with 11 selected Skyrme forces was performed. In Fig. 3 of that
paper, the ratio between the elliptic flow parameter of free neutrons and
protons was plotted as a function of the slope parameter L for Au+Au collision
from 400 to 1000 AMeV. The highest sensitivity was there obtained at 400 AMeV,
while sensitivity at 1000 AMeV was reduced by a factor 2.

Sensitivity of the Au+Au systems is similar to the one of the other neutron rich
systems, even in the case of $^{132}$Sn radioactive ion beams. Heavier systems,
Pb and Au, present smaller statistical errors, for the same numbers of simulated
events. It is also important to stress the differences in trends (slopes) observed in
left panel of Fig.~\ref{fig:fig7}. For the soft EoS the ratios increase with the
energy while for the stiff EoS the trend is opposite.  This proves the needs for
measuring the excitation functions of these observables and the importance of
using neutron rich beams where the effect is stronger. 
We have verified that by filtering the simulations for the acceptance of the common
neutron and proton detector (NeuLAND, see below) does not change the results shown
for the sensitivity of the proposed observable.

\begin{figure}[ht!]
 \begin{minipage}[t]{0.5\linewidth}
 \begin{center}
\includegraphics[width=0.85\textwidth]{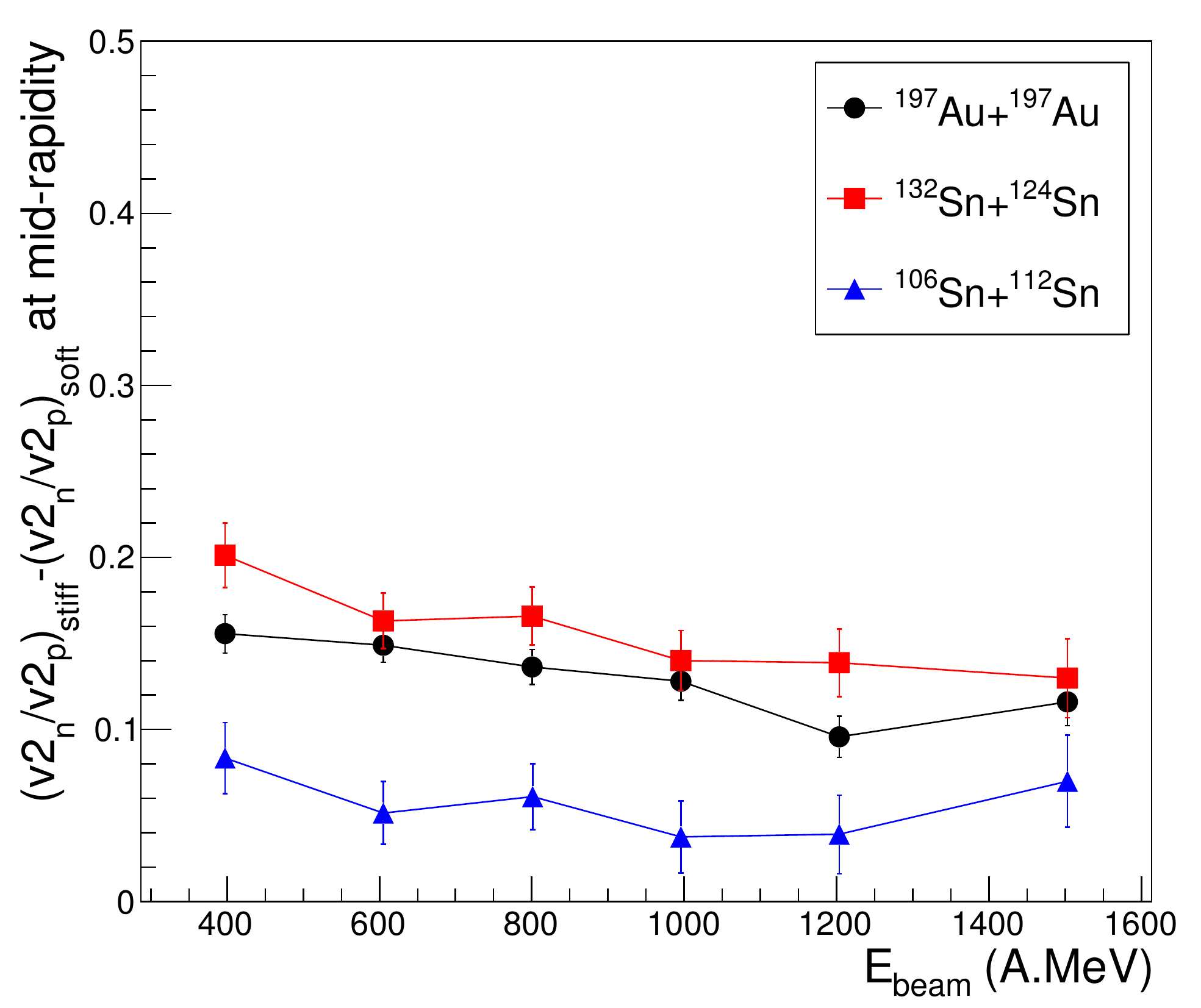}
 \end{center}
 \end{minipage}
 \hfill
 \begin{minipage}[t]{0.5\linewidth}
    \vspace*{-65mm}
 \begin{center}
\includegraphics[width=0.75\textwidth]{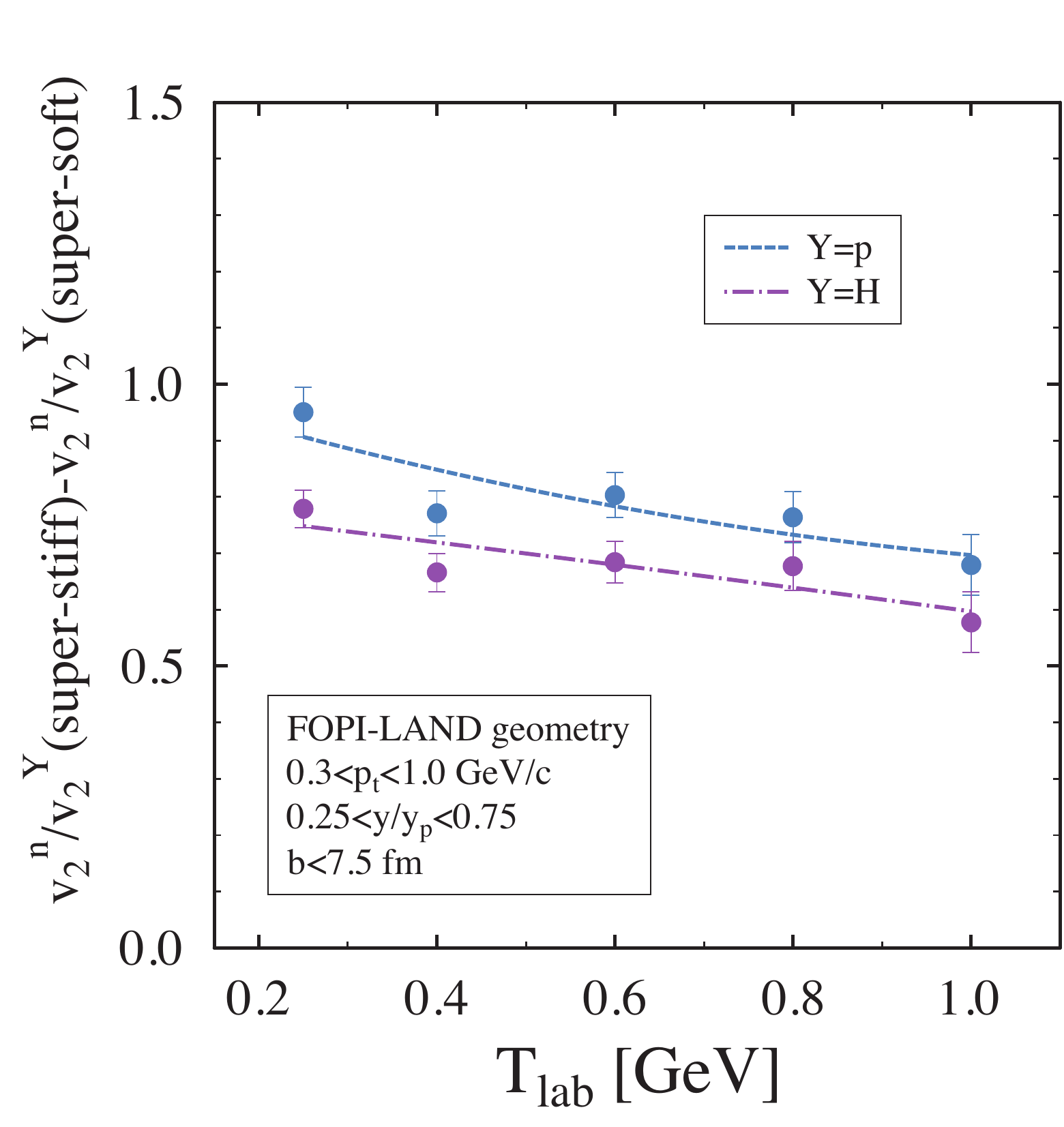}
 \end{center}
 \end{minipage}
\vspace{-3mm}

\caption{Left panel: same as right panel of Fig.~\ref{fig:fig7} but with the
predictions of the IQMD model \cite{ref:hartnack} at b=6 fm. Right panel:
Similar to right panel of Fig.~\ref{fig:fig7} with the predictions of the
T\"uQMD model \cite{ref:cozma2011aa} for mid-central collisions ($b<7.5 fm$)  of
the Au+Au system. The magenta curve provides in addition the flow ratio
obtained  when taking all hydrogen isotopes instead of protons.}
\label{fig:figIQMD1}
\end{figure}

In order to convince ourselves that this level of sensitivity to the symmetry
energy is not just peculiar to the UrQMD approach, we have performed simulations
with other transport models, among them with the IQMD \cite{ref:hartnack}. We show predictions in left panel of Fig.~\ref{fig:figIQMD1} for the same systems as
in Fig.~\ref{fig:fig7}. Comparing with UrQMD, we observe qualitatively similar
trends, except for a higher predicted sensitivity for the highest incident
energy. The right  panel of Fig.~\ref{fig:figIQMD1} shows the predictions of the
T\"uQMD transport model \cite{ref:cozma2011aa} which have been  performed \textit{within
the acceptance cut of the FOPI-LAND experiment, but using two extreme cases for
the $E_{sym}$ parametrizations}, i.e. the x=-2 super-stiff and x=2 super-soft
cases. Also in this case we find a decreasing sensitivity with the increasing
beam energy, with a strong sensitivity of the flow ratios even at the highest
beam energies between the two extreme super-stiff and super-soft choices. Note
here that when using hydrogen isotopes instead of protons, this sensitivity is
reduced. This emphasizes again the necessity to experimentally separate
the hydrogen isotopes.

\vspace{-1mm}

Measuring the excitation functions will provide an additional constraint on
$E_{sym}(\rho)$ through the trends. Availability of the high resolution neutron
detector with a capability to resolve LCPs and relativistic stable heavy ions
makes the upcoming FAIR-Phase-0 facility a unique place to perform proposed
measurements. In the future, once the new R3B-cave will be available, we plan to
pursue this kind of studies by profiting from the  radioactive heavy ions and
detectors there installed.\\

\begin{figure}[h!]
 \centering
 \includegraphics[width=0.7\textwidth]{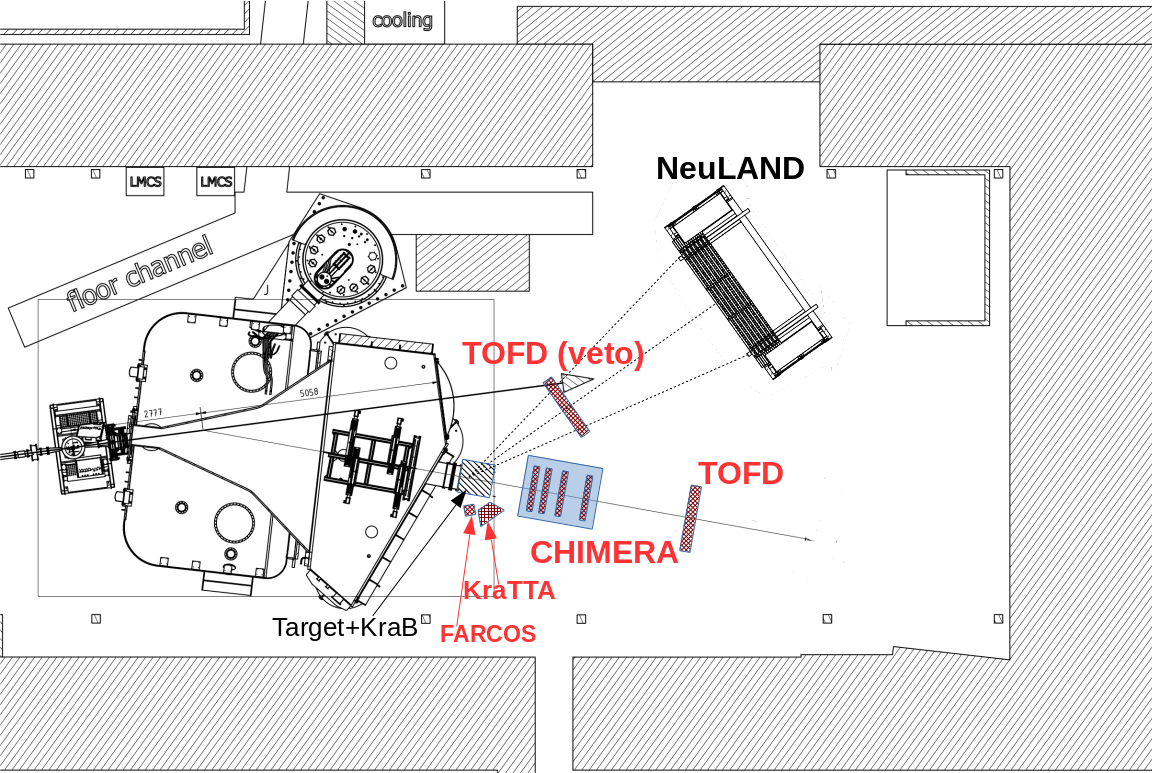}
 \caption{Scale drawing of the proposed set-up for the  experiment in cave C. }
 \label{fig:setup2020}
\end{figure}

\noindent\textbf{\scshape Experimental Design, Methods, Set-Up, Technical
Requirements}  The set-up that we propose to use in the
experiment is sketched in  Fig.~\ref{fig:setup2020}. Eight rings of CsI(Tl) of
CHIMERA \cite{ref:chimera}, covering the polar angles between 7$^{\circ}$ and
20$^{\circ}$ will permit the measurement of charge and energy of forward emitted
LCPs, up to Z $\sim$ 4. These rings fulfill the requirements of both high
granularity, 352 independent modules, and the cylindrical symmetry around the
beam axis. This configuration was already used in the ASY-EOS experiment and
proved to be adequate in order to determine the impact parameter and the
reaction plane orientation of the events \cite{ref:russotto2016}.

Charged particles emitted between $\sim$1$^{\circ}$ and 7$^{\circ}$ 
will be detected by the R3B TOFD. The TOFD consists of two frames. Each frame
has two planes of 44 scintillator paddles with the dimensions
$1000\times27\times5~\mbox{mm}^{3}$. The paddles of successive planes are
shifted by half a paddle width. TOFD will be able to measure charge and velocity
of particles emitted at small angles, mainly projectile spectator-residue and
its decay products. This will be useful to measure global variables for
characterizing the collision centrality, such as $Z_{max}$, charge of the biggest
fragment, and $Z_{Bound}$, amount of charge in clusterized fragments; the TOFD
will serve the same purpose as the Aladin ToFWall in the ASY-EOS experiment.

\begin{figure}[ht!]
 \begin{minipage}[t]{0.5\linewidth}
 \begin{center}
\includegraphics[width=0.95\textwidth]{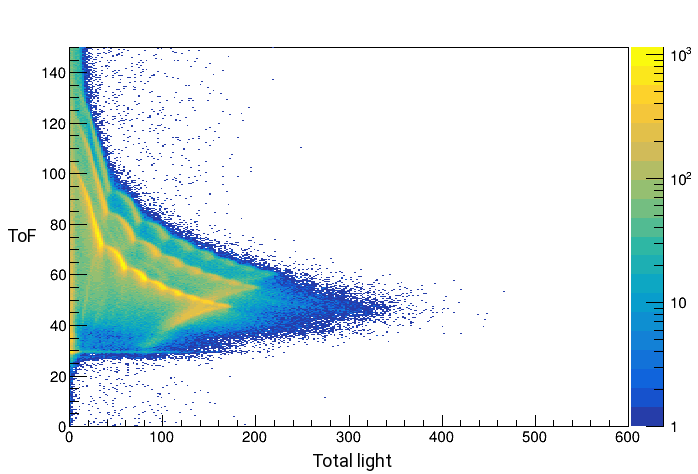}
 \end{center}
 \end{minipage}
 \hfill
 \begin{minipage}[t]{0.5\linewidth}
    \vspace*{-55mm}
 \begin{center}
\includegraphics[width=0.95\textwidth]{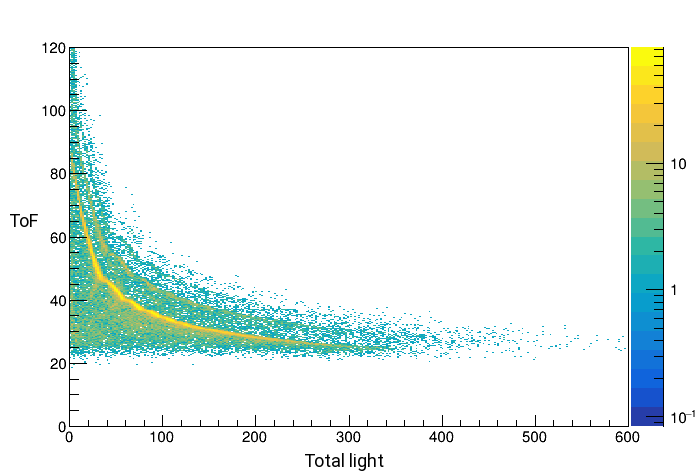}
 \end{center}
 \end{minipage}
\vspace{-7mm}

\caption{Left panel: Time-of-flight (in ns) vs Total light deposit in the
NeuLAND demonstrator as obtained from Sn+Sn @ 270 AMeV measurement at RIKEN;
Right panel: Time-of-flight vs Total light collected in 12 double-planes of
NeuLAND from the simulation of Au+Au reaction at 400 AMeV (by I. Gasparic).}

\label{fig:neuland}
\end{figure}

The NeuLAND detector \cite{ref:NeuLAND} will give a unique opportunity to
measure the neutron and LCP in the same angular regions. The outstanding
calorimetric properties of NeuLAND will allow  protons and other hydrogen
isotopes to be relatively well separated, and will give access to the 
neutron vs proton observables. The NeuLAND demonstrator was a part of the SPiRIT
experiment \cite{ref:Spirit} carried out at RIKEN in 2016 and the capability of
resolving both protons and neutrons was clearly demonstrated there. The
identification plot of hydrogen isotopes in the demonstrator (4 double planes,
40 cm total thickness) is presented in the left panel of Fig. \ref{fig:neuland}.
The p, d, t lines are clearly resolved up to the punch through energy (about 260
MeV for protons) above which the characteristic back-bendings occur.


The foreseen 12 double planes of NeuLAND, resulting in a total depth of 120 cm,
will assure stopping of protons up to about 500 MeV. A simulated identification
plot for the Au+Au collisions at 400 AMeV is presented in the right panel of
Fig. \ref{fig:neuland}. Indeed, no punch-through segments are observed at this
energy and the p, d, t lines clearly stick out of the secondary reaction and
multi-hit background. The simulations include tracking, the secondary reaction
losses, multiple Coulomb scattering, light propagation in plastic scintillators
and quenching effects. The estimated efficiency for proton identification
amounts to about 64\% at 200 MeV and 36\% at 400 MeV. Taking into account the
thickness of the NeuLAND calorimeter and the secondary reaction and scattering
probability, the estimated efficiencies are still impressive. The identification
capability within the punch-through segments at higher energies can be improved
by applying statistical methods including regularized decomposition as shown in
\cite{ref:ockham}.

With NeuLAND in its start-up version, consisting of 120 cm detector depth (12 double planes), the one neutron interaction probability is about 70\% at 400 MeV \cite{ref:NeuLAND_tech}. Taking into account also the reconstruction efficiency a five-neutron event is recognized with correct neutron multiplicity with a probability of about 20 to 30\% (200 to 1000 MeV). 


NeuLAND will be placed 5.8 m away from the target covering effectively the
mid-rapidity regions, i.e., the polar angles between 33$^{\circ}$ and
57$^{\circ}$. In order to better discriminate neutrons from protons/LCP a second
TOFD will be placed at a distance of 2 m from the target, in geometrical
correspondence with the NeuLAND, playing a role of the veto wall.




\begin{figure}[ht!]
 \begin{minipage}[t]{0.55\linewidth}
 \begin{center}
\includegraphics[width=0.89\textwidth]{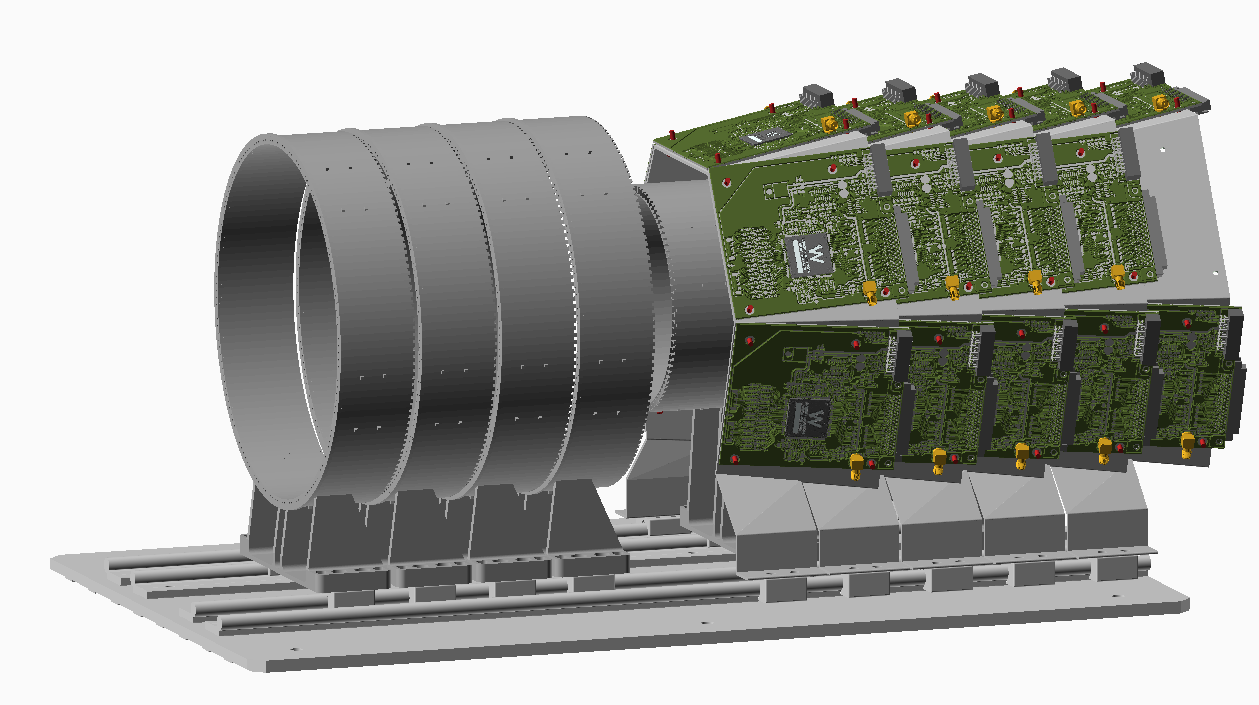}
 \end{center}
 \end{minipage}
 \hfill
 \begin{minipage}[t]{0.44\linewidth}
    \vspace*{-48mm}
 \begin{center}
\includegraphics[width=0.95\textwidth]{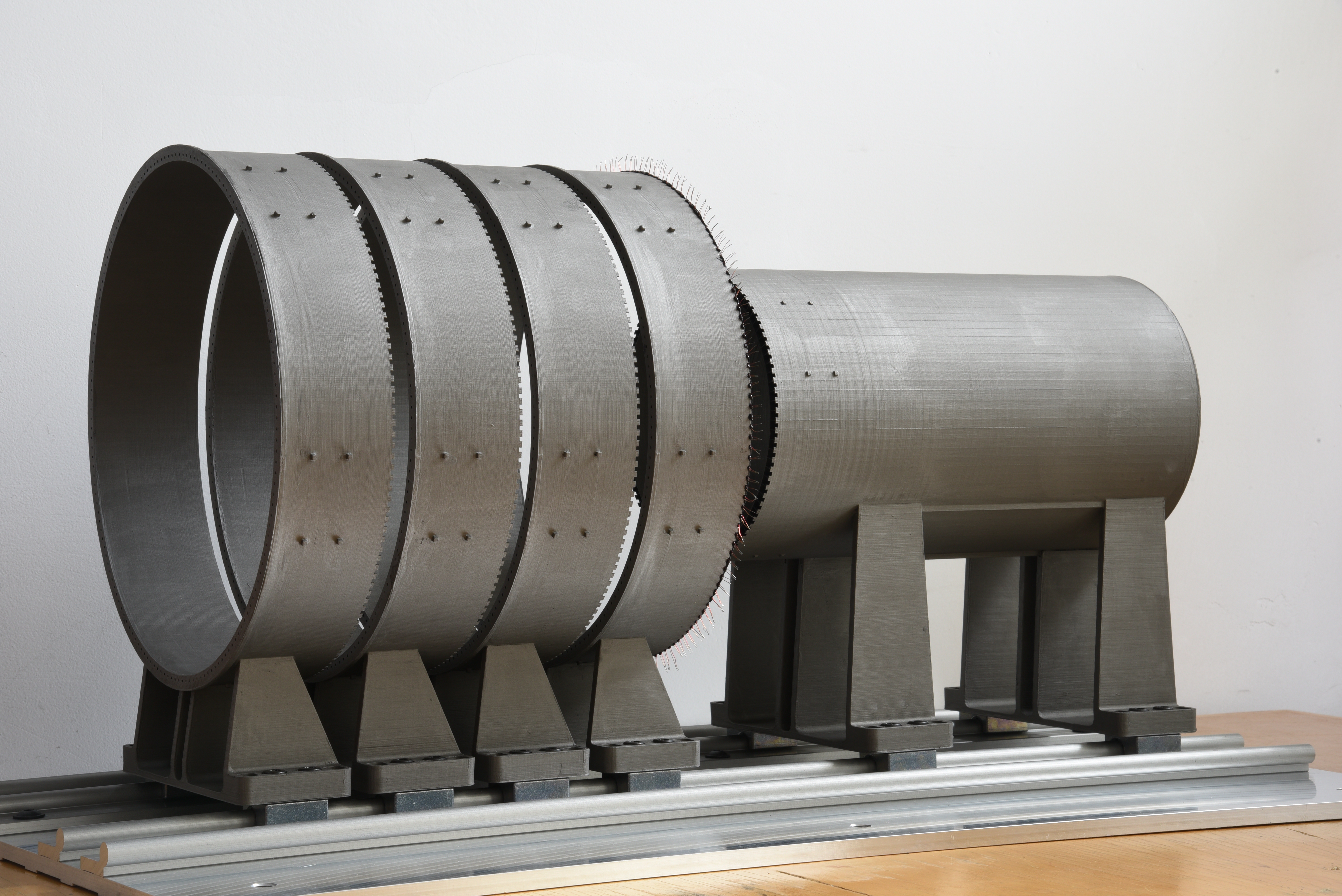}
 \end{center}
 \end{minipage}

\caption{Left panel: The design of the KRAB detector including the CITIROC
boards; Right panel: The actual view of the KRAB as of the end of March 2020.}

\label{fig:krabbing}
\end{figure}

The KRAB (KRAkow Barrel) detector \cite{ref:jl_krab} has been designed to
provide a fast trigger signal based on the multiplicity threshold as well as
precise azimuthal distributions for charged particles beyond the angular
acceptance of the CHIMERA+TOFD setup. The main features of the KRAB detector
are: 5 rings of 4$\times$4 mm$^{2}$ fast scintillating fibers (BCF-10, SciFi)
read out by the MicroFJ-30035-TSV SiPMs, covering polar angles from $30^{\circ}$
to $165^{\circ}$ with $\sim$87\% geometrical efficiency and with $\sim$5\%
multi-hit probability. It will be sufficiently large for radioactive beams and
sufficiently small and lightweight in order not to disturb neutrons, having the
min and max internal radii of 6.9 and 11.5 cm and a length of $\sim$50 cm. It
will consist of 4$\times$160 segments in forward rings and 96 segments in the
backward ring with a total of 736 channels. The mechanical structure holding the
SciFi segments and the front end electronics has been 3D-printed with the ABS
filament. The SiPMs will be read out and controlled using the 32 channel CITIROC
ASICs. In the ASY-EOS experiment the backward region was covered by the
MicroBall detector consisting of 50 CsI(Tl) crystals arranged in 4 rings. It was
used to roughly define the reaction-plane orientation in the backward region. In
addition the correlation between the impact vectors deduced from the backward
and forward azimuthal distributions measured by the MicroBall and CHIMERA
detectors was found to be of fundamental importance to efficiently reject the
off-target upstream reactions in air. Thanks to its high segmentation, the KRAB
detector will allow to measure precisely the azimuthal distributions which are
indispensable for high resolution estimates of the reaction plane. It will also
produce a fast trigger based on total multiplicity in an angular region of
$\theta>30^{\circ}$ where a strong correlation between the multiplicity and the
magnitude of the impact parameter is expected from the model predictions. This
will allow for more precise centrality estimates than with the CHIMERA detector
alone.  Moreover, the design of KRAB assumes construction of a ``helium
sleeve'', with the target holder inside. Simulations indicate that this should
reduce the number of unwanted hits caused by delta-electrons by a factor of
about 30. Thus, it is expected that KRAB should greatly improve the quality of
the data with respect to the first ASY-EOS experiment.  The design and the
actual view of the KRAB detector are presented in Fig. \ref{fig:krabbing}. 

In addition, yields and flows of LCPs at mid-rapidity and at backward angles
will be measured by using the KRATTA triple telescope array \cite{ref:lukasik}
and the Femtoscope Array for Correlations and Spectroscopy (FARCOS)
\cite{ref:FARCOS}, respectively. Measuring precisely yields and isotopic
compositions of clusters emitted at mid-rapidity and from the target-spectator
is of fundamental importance for advanced tuning of clusterization algorithms
used in the transport models to get more realistic predictions from them. In
fact clustering influences also the predictions of neutron/proton yields and
flows, with clusters acting as absorbers of the otherwise free nucleons.
Moreover, transport models predict a sensitivity of the ratio of  yields and
flows of light isobar nuclei to the high density behavior of the $E_{sym}$
\cite{ref:Gio10}, and measuring isotopic composition of clusters will allow the
use of thermodynamic methods (thermometry) to study properties of the emitting
sources, such as their temperature. The high angular resolution of FARCOS,
$\sim$0.25$^{\circ}$ will perfectly suit the measurements of particle-particle
correlation functions \cite{ref:Ser98} and, through interferometry, will allow
for characterization of space-time properties of emitting sources.\\


\noindent\textbf{\scshape Justification of Beamtime Request.}
The system/energies we want to measure in the proposed campaign are:
\begin{center}
\begin{tabular}[h]{ccl}
$^{197}$\textnormal{Au} + $^{197}$\textnormal{Au} &at& 250, 400, 600, 1000 AMeV\\
\end{tabular}
\end{center}

Motivation of the 4 energies is the following:
\begin{itemize}

\item the 250 AMeV energy, according to calculations shown in Fig. 13 of
\cite{ref:coz18}, is the energy showing the highest sensitivity on $K_{sym}$;
this energy corresponds to the dynamic range of KRATTA assuring clean p, d, t
identification without punch-through segments;

\item the 400 AMeV energy is the one measured in the past ASY-EOS experiment and
is necessary as a reference point, capable of unveiling systematic difference
between the new and old measurements; this is the energy of the maximum
squeeze-out, it also assures no punch-through segments in NeuLAND identification
maps;

\item the 600 AMeV energy, according to calculations shown in Fig. 13 of
\cite{ref:coz18}, is the energy showing the highest sensitivity on L;

\item the 1000 AMeV energy is the energy allowing to explore the highest
densities where the neutron/proton elliptic flow observable keeps a significant
sensitivity ($\sim15\%$) to symmetry energy parametrization, according to the UrQMD
calculations of Fig.~\ref{fig:fig7};

\end{itemize}

\noindent Let us stress again the importance of measuring all these energies to
provide additional constraint on the Symmetry Enegy through the observed trends,
which are expected to be opposite for the soft and stiff assumptions (see
Fig.~\ref{fig:fig7}).

\noindent\textbf{\scshape Acknowledgement:}\\
Work supported by Polish National Science Centre, contract No. UMO-2017/25/B/ST2/02550.

\end{document}